\documentclass[%
 amsmath,amssymb,
 aps,
prl,
 reprint,
showpacs
,superscriptaddress,groupedaddress
]{revtex4-1}
\usepackage{epsfig}
\usepackage{hyperref}
\input epsf.tex
\usepackage{amssymb}
\usepackage{amsmath}
\usepackage{amsthm}
\usepackage{amsopn}

\def\comment#1{}
\def\bra#1{\mathinner{\langle{#1}|}}
\def\ket#1{\mathinner{|{#1}\rangle}}

\def\beq{\begin{equation}}
\def\eeq{\end{equation}}
\def\bea{\begin{eqnarray}}
\def\eea{\end{eqnarray}}

\usepackage{ulem}
\usepackage{cancel}
\usepackage{color}

\hypersetup{
    colorlinks=true,       
    linkcolor=blue,          
    citecolor=blue,        
    filecolor=magenta,      
    urlcolor=blue           
}
\begin{document}

\title{Probability Density of Relativistic Spinless Particles }

\author{M. J. Kazemi}

\email{mj_kazemi@sbu.ac.ir}
\author{H. Hashamipour}
\email{h_hashamipour@sbu.ac.ir}
\affiliation{ Department of Physics, Faculty of Science, Shahid Beheshti University, Tehran 19839, Iran}
\author{M. H. Barati}
\email{mohbarati14@gmail.com}
\affiliation{Department of Physics, Faculty of Science, Kharazmi University, Tehran, Iran}

\begin{abstract}
In this paper, a new conserved current for Klein-Gordon equation is derived. It is shown, for $1+1$-dimensions, the first component of this current is non-negative and reduces to $|\phi|^2$ in non-relativistic limit. Therefore, it can be interpreted as the probability density of spinless particles. In addition, main issues pertaining to localization in relativistic quantum theory are discussed, with a demonstration on how this definition of probability density can overcome such obstacles. Our numerical study indicates that the probability density deviates significantly from $|\phi|^2$ only when the uncertainty in momentum is greater than $m_0c$.


\end{abstract}

\maketitle

\section{Introduction}
The Born's interpretation of $|\psi|^2$ as position probability density  is one of the most fundamental axioms of quantum mechanics as it  provides a link between the mathematical formalism and empirical results \cite{M. Born}. This axiom has been incredibly successful in predicting position probability density in non-relativistic quantum mechanics. Although, in the relativistic regime, the quadratic relation between position probability density and wave function has been confirmed by recent high accuracy single photon multi-slit experiments \cite{Sinha 2010, Hickmann 2011, Gagnonn 2014, Sawant 2014, Kauten 2017}, a satisfactory  mathematical expression for position probability density of relativistic bosons has not yet been found. In the simplest case, finding a well-defined position probability density for the free spinless particles is a long-standing problem (see, e.g., \cite{Weinberg 1995, Nikolic 2007}): The time component of the well-known Klein-Gordon conserved current, ${J_{KG}^\mu}{=}i({\phi}^{*}\partial^\mu \phi -\phi {\partial^\mu \phi}^{*})$, may be negative on some regions of space-time and can not be interpreted as position probability density \cite{footnote1}. One may suggest to use the $|\phi|^2$ as probability density, similar to the non-relativistic theory \cite{Kowalski, Rembielinski, Horwitz 1985}; In which case it is easy to see that the Klein-Gordon equation,
\begin{equation}\label{1}
\Box \phi+m^2\phi=0,
\end{equation}
 leads to the following continuity equation for $|\phi|^2$  \cite{Kowalski}: 
\begin{equation}\label{2}
\partial_t\rho_B+\nabla.\textit{\textbf{J}}_{B}=0,
\end{equation}
where 
\begin{equation}\label{3}
\rho_B=|\phi(x)|^2=N\int  \tilde{\phi}(p) \tilde{\phi}^{*}(k) e^{i(p-k).x}\ d^4p\ d^4k,
\end{equation}
\begin{equation}\label{4}
\textit{\textbf{J}}_{B}=N\int \tilde{\phi}(p) \tilde{\phi}^{*}(k)\ e^{i(p-k).x}\ \textit{\textbf{u}}(p,k)\ d^4p\ d^4k, 
\end{equation}
\begin{equation}\label{5}
\textit{\textbf{u}}(p,k)=\frac {\textit{\textbf{p}}+\textit{\textbf{k}}  }{p_0+k_0},
\end{equation}
and $\tilde\phi(p)$ is Fourier-transformation of wave function.
It should be noted, despite the fact that $|\phi|^2$ is non-negative and conserved, it cannot be considered as position probability density: due to the Lorentz length contraction, the probability density can not be a scalar \cite{Sakurai}. In other words, $J_B^\mu=(\rho_B,\textit{\textbf{J}}_{B})$ is not a four-vector and therefore cannot be interpreted as a relativistic probability current density \cite{Horwitz 1985}. In addition, Born's probability density leads to faster than light particle propagation \cite{Peskin 1995, Padmanabhan 2016}. In principle, a reasonable probability current must satisfy the following conditions: 
\\

I)  Lorentz transformation: \ $J^{\mu'}=\Lambda_\mu^{\mu'} J^\mu $,
\\

II) probability conservation: \ $\partial_\mu J^\mu=0$,
\\

III) future-orientation: \ $J^0\geq 0$,
\\

IV) causal propagation: \ $J^\mu J_\mu\geq 0$.
\\

The last condition is necessary as it ensures the causal propagation of particles. In fact, there are several other currents which have been suggested for the Klein-Gordon equation \cite{Mostafazadeh 2006, E. Marx 1972} all of which do not satisfy, at least, one of the above conditions. The aim of this paper is to propose a proper expression for relativistic probability current that satisfies all of the aforementioned conditions.

\section{Position Distribution}
 According to equations (\ref{3}) and (\ref{4}), we suggest the following expression as the relativistic probability current \cite{footnote2}:
\begin{equation}\label{6}
J^\mu{=}\int \tilde{\phi}(p) \ \tilde{\phi}^{*}(k)\ e^{i(p-k).x}\ u^\mu(p,k)\ d^4p\ d^4k,
\end{equation}
where $u^\mu(p,k)$ is an unknown function that must be determined by theoretical constrains. In this regard, the condition (I) implies that the $u^\mu(p,k)$ is a four-vector. The general form of a four-vector made by $p$ and $k$ is given by    
\begin{equation}\label{7}
u^{\mu}(p,k)=\alpha (p^\mu+k^\mu) + \beta (p^\mu-k^\mu),
\end{equation}
where $\alpha$ and $\beta$ are scalar coefficients. Next, the conservation condition (II) leads to $\beta =0$. Also, in principle, the coefficient $\alpha$ should be determined using conditions (III) and (IV). This procedure in (1 + 1)-dimensions is straightforward and a possible choice is (see appendix A)
\begin{equation}\label{8}
\alpha(p,k)=  \frac{ \xi}{\sqrt{(p+k)^2}},
\end{equation}
where $\xi=\frac{1}{2}(\frac{k^0}{|k^0|}+\frac{p^0}{|p^0|})$. 
So finally we get the  $u^{\mu}(p,k)$ as follows:
\begin{equation}\label{9}
u^{\mu}(p,k)= \xi\frac{p^{\mu}+k^{\mu}}{\sqrt{(p+k)^2}}.
\end{equation}
Equation (\ref{9}) can be rewritten as $u^{\mu}=|\xi|\gamma(1,\textit{\textbf{u}})$, in which $\textit{\textbf{u}}$ is the velocity-vector  defined in equation (\ref{5}) and $\gamma=1/\sqrt{1-{\textit{\textbf{u}}}^2}$ is the corresponding Lorentz coefficient. In fact, the expression (\ref{6}) is the simplest covariant generalization of the equations (\ref{3}) and (\ref{4}). The only difference between this expression and  the Born probability current, $J_B^\mu$, is the factor $|\xi|\gamma$: 
\begin{equation}\label{10}
\rho(x)=N\int |\xi|\gamma \tilde\phi(p) \tilde\phi^{*}(k)\ e^{i(p-k).x}\ d^4p\ d^4k,
\end{equation}
\begin{equation}\label{11}
\textit{\textbf{J}}(x)=N\int |\xi|\gamma \textit{\textbf{u}}\ \tilde\phi(p) \tilde\phi^{*}(k) e^{i(p-k).x}\ d^4p\ d^4k,
\end{equation}
The factor $\gamma$  comes naturally in accordance with Lorentz contraction and the factor $|\xi|$ prohibits the occurrence of \textit{Zitterbewegung} behavior \cite{Schrodinger, Krekora}. In appendix A, it is shown that, for massive particles in $(1+1)$-dimensions, equations (\ref{10}) and (\ref{11}) can be rewritten in position representation as follows:
\begin{equation}\label{12}
\rho=| {\mathcal{D}}^+ \phi_+|^{2}+| {\mathcal{D}^-}\phi_+|^{2}+|{\mathcal{D}^+}\phi_- |^{2}+| {\mathcal{D}^-}\phi_- |^{2},
\end{equation}
\begin{equation}\label{13}
J=\left(| {\mathcal{D}}^+ \phi_+|^{2}-| {\mathcal{D}^-}\phi_+|^{2}+|{\mathcal{D}^+}\phi_- |^{2}-| {\mathcal{D}^-}\phi_- |^{2}\right) c,
\end{equation}
where $\phi_{\pm}$ are positive and negative frequency components of wave function, $\phi=\phi_{+}+\phi_-$, and ${\mathcal{D}}^{\pm}$ are pseudo-differential operators which are defined as follows:
\begin{equation}\label{14}
{\mathcal{D}^{\pm}}\equiv \sqrt{\frac{1}{2} (\sqrt{1-\lambda_c^2\frac{d^2}{dx^2}}{\mp}i\lambda_c\frac{d}{dx})},  
\end{equation}
in which $\lambda_c\equiv\hbar/mc$ is Compton wave length. From equations (\ref{12}) and (\ref{13}) It is clear that, the probability density is unambiguously positive definite and $|J/{\rho}|\leq c$.

 It is clear that when the wave function only has positive energy part \cite{Newton wave funcion}, $\phi=\phi_+$, Klein Gordon equation leads to 
 \begin{equation}\label{15}
i\hbar \frac{\partial  \phi}{\partial t} =\sqrt{-\nabla^2+m^2} \phi. 
\end{equation}
and equations (\ref{12}) and (\ref{13}) reduce to the following simpler forms:
\begin{equation}\label{16}
\rho=| {\mathcal{D}}^+ \phi|^{2}+| {\mathcal{D}^-}\phi|^{2},
\end{equation}
\begin{equation}\label{17}
J=\left(| {\mathcal{D}}^+ \phi|^{2}-| {\mathcal{D}^-}\phi|^{2}\right) c.
\end{equation}
In this case, in non-relativistic regime ($c\to \infty$), the equation (\ref{15}) reduces to non-relativistic Schr\"{o}dinger equation, also  equations (\ref{16}) and (\ref{17}) reduce to non-relativistic probability density, $|\phi|^2$, and conventional Schr\"{o}dinger probability current, $(\hbar/m) \Im({\phi}^*\partial_x\phi)$, respectively.

\begin{figure}[t]
	\includegraphics[width=0.45\textwidth, height=0.60\textwidth, trim={0cm 4cm 0cm 4cm}]{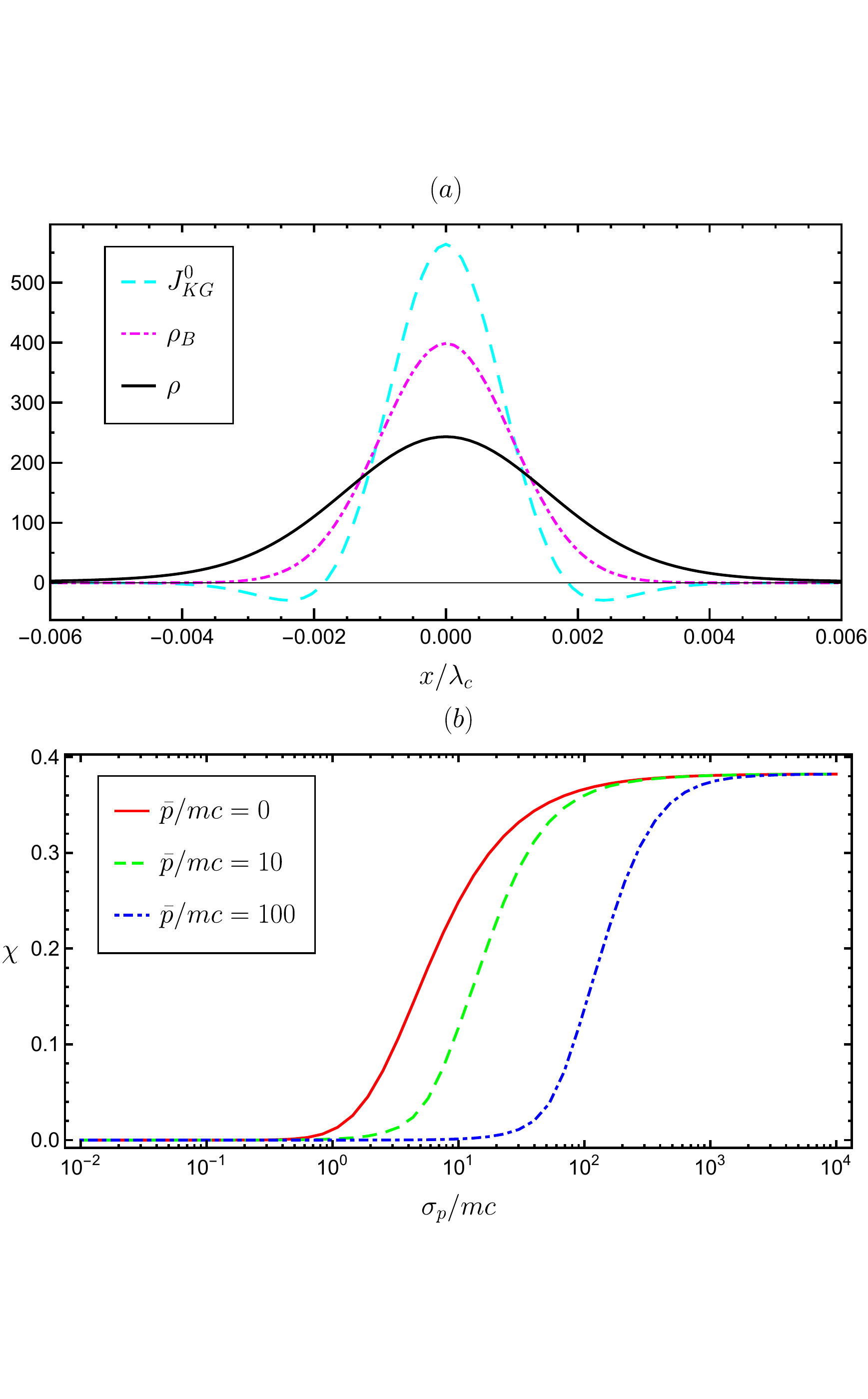}
	\caption{\label{fig:epsart}(a) The first component of Klein-Gordon current  $J^0_{KG}$ (dashed line), the Born probability density $\rho_B$ (dash-dotted line) and the relativistic probability density $\rho$ (solid line) referring to the Gaussian wave function (\ref{19}) with $\sigma_p/mc=1000$ and $\bar{p}/mc=0$. (b) Represents the $\chi$  for Gaussian wave function (\ref{19}).}\label{fig1}
\end{figure}

For comparing the relativistic probability density (\ref{16}) with $|\phi|^2$, In Figure  \ref{fig1}, we plot $\chi$ (a measure of deviation from Born probability) defined as: 
\begin{equation}\label{18}
\chi=\int_{-\infty}^\infty \left|\rho-|\phi|^2\right|\ dx,
\end{equation}
for this Gaussian wave function
\begin{equation}\label{19}
\tilde{\phi}(p)=N e^{-(p-\bar{p})^2/\sigma_p^2}.
\end{equation}
From figure (1-b) it is clear that, when momentum uncertainty is small compared to $mc$, relativistic probability density deviation from  Born probability density is negligible, even assuming that the group velocity of wave packet is comparable with velocity of light.

Finally, it should be noted, although the expression for probability density in terms of wave function, equation (\ref{16}), is non-local, there is no inconsistency with special relativity. In fact, this non-locality is  essential to introduce a self-consistent relativistic probability density; since the relativistic wave function  can propagate outside the light cone, a local relation between wave function and  probability density, for instance $\rho=|\phi|^2$, leads to faster than light particles propagation \cite{Padmanabhan 2016,Peskin 1995}. In the following section, the relativistic requirements imposed on the definition of probability density is further discussed together with an account of how our suggested expression satisfy them.

\section{localization and Causality}

It is well-known there are some problems on  concept of "localized particle" in relativistic quantum mechanics \cite{Foldy Wouthuysen 1950, Newton 1949, Rosenstein 1981, Rosenstein 1987, Thaller 1992, Hegerfeldt 1974, Skagerstam 1976, Perez 1977, Hegerfeldt 1980, Hegerfeldt 1985, Bracken 1999, Bracken 2005, Bracken 2007}. The notion of localization is closely related to the concept of position probability density. In this section we will briefly review these problems and demonstrate how our definition of relativistic probability density can circumvent such obstacles.

One of the earliest attempts to analyze the notion of localized particle in relativistic quantum mechanics was made by Newton and Wigner. In 1949 they uniquely derived a relativistic position operator and its eigenstates using some justifiable postulates about the exact localized states \cite{Newton 1949}. However, the Newton-Wigner position operator, although arising from seemingly reasonable postulates, suffers from the following drawbacks:
\begin{itemize}
  \item	A state which is exactly localized in one reference frame, i.e. an eigenstate of Newton-Wigner position operator, is not localized in other reference frames \cite{Newton 1949}.
  \item The definition of position probability density based on the Newton-Wigner position operator, i.e. $\rho_{NW}=|\psi_{NW}|^2$ \cite{Newton wave funcion}, leads to faster than light particle propagation \cite{Rosenstein 1981, Rosenstein 1987, Thaller 1992}.
\end{itemize}
These difficulties indicate that the Newton-Wigner position operator is not quite acceptable. Moreover, it has been shown that, for a general case, any strict localization leads to superluminal propagation \citep{Hegerfeldt 1974, Skagerstam 1976, Perez 1977, Hegerfeldt 1980, Hegerfeldt 1985}. An apparent way out of this problem is to assume that such strict localization is not possible. This implies that a proper relativistic self-adjoint position operator does not exist \cite{Thaller 1992, Hegerfeldt 1985}, and hence defining the position distribution via the projection valued-measure associated to position operator is not realizable  \cite{note}. A possible treatment is to introduce a reasonable probability density without recourse to a position operator \cite{Hegerfeldt 1974}, as the one presented in this paper. It must be emphasized, the problem of superluminal propagation is not just the characteristic behavior of Newton-Wigner's probability density. Hegerfeldt proved    \cite{Hegerfeldt 1974, Skagerstam 1976, Hegerfeldt 1980}, on very general grounds and for any reasonable definition of probability density, that a particle initially localized with probability 1 in a finite volume of space, immediately develops infinite "tails". In what follows, we prove a theorem that shows how our probability density keeps the particle from strict localization, which is  the main requirement of Hegerfeldt theorem. This is similar to what Thaller proved for the case of Dirac probability density \cite{Thaller 1992}. 

\textbf{Theorem.} Let $\rho$ be the probability density associated with an arbitrary positive-energy wave function $\phi$, presented in equation(\ref{16}). Then
\begin{equation}\label{20}
\textnormal{Supp}(\rho)=\mathbb{R}	
\end{equation}
where $\textnormal{Supp}(\rho)$ stands for support of $\rho$ which is defined as
\begin{center}
$\textnormal{Supp}(\rho)\equiv \textnormal{Closure  of} \ \{x\in \mathbb{R} \ | \ \rho(x)\neq 0 \}.$
\end{center}

\textbf{Proof.} From equation (\ref{16}) it is clear, for a particle to be strictly localized in a compact subset of $\mathbb{R}$, the supports of $\mathcal{D}^+\phi$ and $\mathcal{D}^-\phi$ should be compact subsets of $\mathbb{R}$. On the other hand, by Paley-Wiener-Schwartz theorem \cite{Paley Wiener 1934, Schwartz 1952}, the Fourier transform of a compactly supported function is guaranteed to be analytic anywhere on the complex plane. But Fourier transform of  $\mathcal{D}^+\phi$ and $\mathcal{D}^-\phi$ cannot be simultaneously analytic; since they are related to each other by
\begin{equation}\label{21}
\widetilde{{\mathcal{D}^+}\phi}(p)=\frac{1}{m}(\sqrt{p^2+m^2}+p)\widetilde{{\mathcal{D}^-}\phi}(p).
\end{equation}
The branch cut in $\sqrt{p^2+m^2}$ at $p=im$ means both $\widetilde{{\mathcal{D}^+}\phi}$ and $\widetilde{{\mathcal{D}^-}\phi}$ cannot be analytic when $p$ is imaginary with magnitude $m$. Hence, this proves the theorem. $\blacksquare$  
\\

The above theorem implies that there is no state for which the probability of finding the particle in a set $\Delta$ is 1 unless $\Delta =\mathbb{R}$. 
Nevertheless, the strict localization of particle is irrelevant for most practical purposes, and it is quite sufficient to adopt an appropriate notion of localization with adjustable precision. It must be emphasized, although  our probability density has tails extending to infinity, arbitrarily small values of position uncertainty is possible.
 In fact for any point of space, $a\in \mathbb{R}$, there is a sequence of wave functions, $\{\phi_n\}_{n=1}^{\infty}$, whose corresponding probability density sequence, $\{\rho_n\}_{n=1}^{\infty}$, approaches $\delta(x-a)$; see appendix B.
This fact indicates that the particle could be localized arbitrarily sharply in the vicinity of any given point. This notion of "arbitrary precise localization" differs from the  one introduced by Newton and Wigner i.e. "exact localization", and was initially employed by Bracken and Melloy for the case of free Dirac electrons \cite{Bracken 1999, Bracken 2005, Bracken 2007}. It naturally avoids the problems plaguing Newton-Wigner's exact localization; Firstly, the localization defined in this sense has correct properties under Lorentz transformations as $J_\mu$ is a covariant vector \cite{Bracken 1999}, secondly, since velocity of probability flow, $J/\rho$, is less than the speed of light, propagation of particle is guaranteed to be causal.


\section{Momentum Distribution}
Since the position probability density deviates from $|\phi(x)|^2$ in relativistic regime, one may raise the question of whether momentum probability distribution deviates from $|\tilde{\phi}(p)|^2$ as well?  To answer this question, we note that, Based on the quantum theory of measurement, each physical measurement can be described as a position measurement: In principle, the variables that account for the outcome of an experiment are ultimately particle positions \cite{Bell, Holland, Durr, Wheeler}. This fact has been made clear by John Bell \cite{Bell}:

 \textit{"In physics the only observations we must consider are position observations, if only the positions of instrument pointers.  ... If you make axioms, rather than definitions and theorems, about the "measurement" of anything else, then you commit redundancy and risk inconsistency."}

\begin{figure}[t]
	\includegraphics[width=0.45\textwidth, height=0.60\textwidth, trim={1cm 0cm 1cm 0cm}]{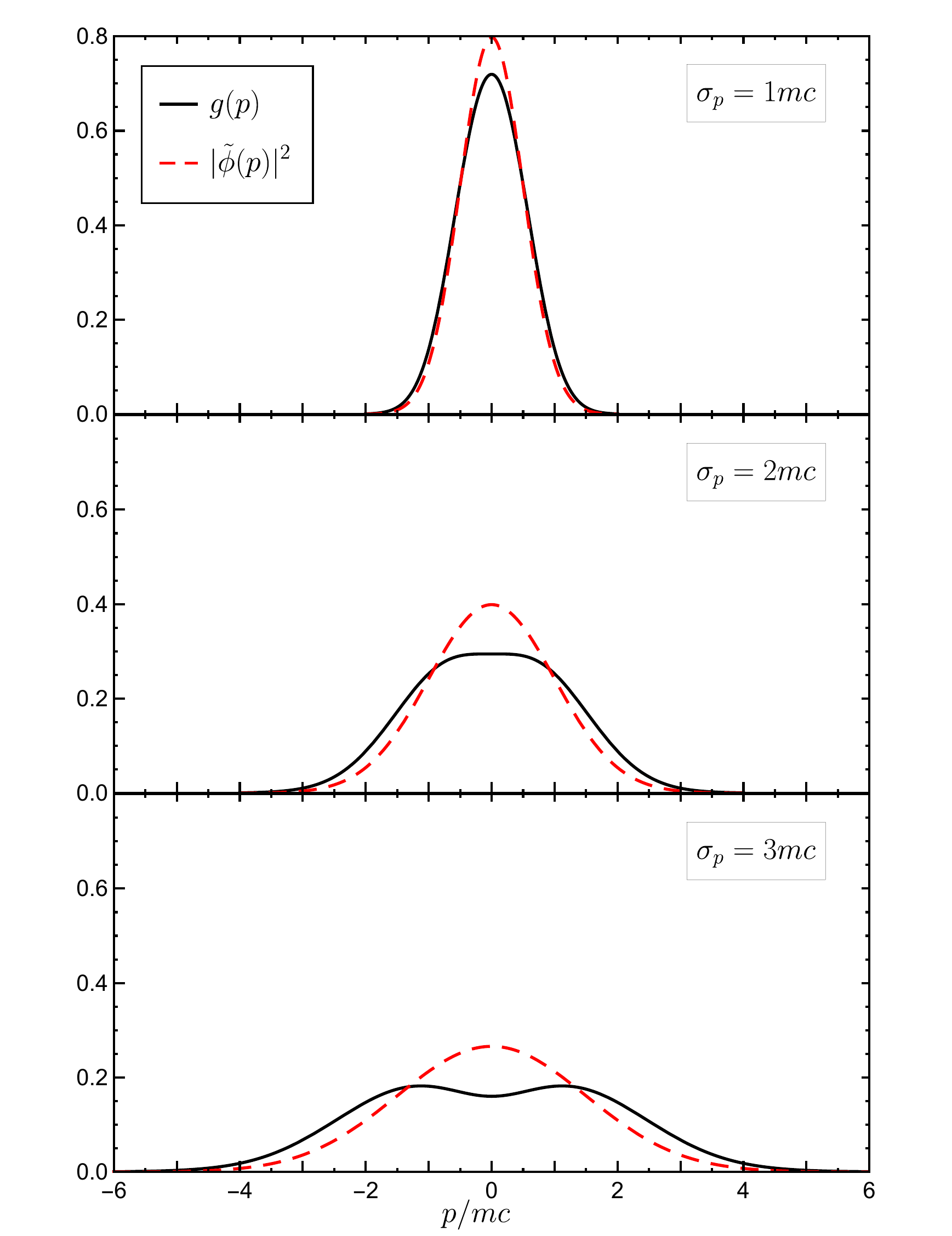}
	\caption{The plot of the relativistic and non-relativistic momentum probability density referring to the Gaussian wave function (\ref{19}) with $\bar{p}=0$.}
	\label{kkk}
\end{figure}

 In this regard, It is shown in non-relativistic quantum mechanics that the Born's rule for any observable can be derived considering the Born's rule on position of particles \cite{Bell, Holland, Durr}. Here we aim to propose a derivation of relativistic momentum probability density from the relativistic position probability density (\ref{16}). The given argument is  based on the Feynman's method for initially confined systems, namely the Time-of-flight measurements \cite{Holland, Feynman 1965, Park 1968, application}. 
 Suppose the wave function is initially confined to a region $\Delta$ centered around the origin $x_0=0$ and is negligible elsewhere. After allowing the wave function to freely propagate for a considerable amount of time, a measurement on the position $x$ of the particle is effected. The probability of the particle's momentum lying inside the element $dp$ around the point $p$ at $t = 0$ is equal to probability of finding the particle's position in the element  $dx$ around the point $x=vt$ provided the limit $t\to \infty$ is taken in order to discard the effect of uncertainty in initial position. So we have:
\begin{equation}\label{22}
g(p)dp=\lim_{t\to\infty}[\rho(x,t)dx]_{x=vt},
\end{equation}
where $g(p)$ represents the momentum probability density and $v=p/E$. using the relativistic position probability density (\ref{16}), the equation (\ref{22}) leads to  
\begin{equation}\label{23}
g(p)=\dfrac{m^2}{E^3}\lim_{t\to\infty}t\left(| {\mathcal{D}}^+ \phi|^{2}+| {\mathcal{D}^-}\phi|^{2}\right)_{x=pt/E}.
\end{equation}
In the non relativistic regime ($c\to \infty$) the equation (\ref{23}) leads to
\begin{equation}\label{24}
g(p)=\dfrac{1}{m}\lim_{t\to\infty}[t|\phi(x,t)|^2]_{x=pt/m}.
\end{equation}
In this case, the Schr\"{o}dinger equation for an initially confined wave function leads to $\phi(pt/m,t)\sim t^{-1/2}\tilde\phi(p)$ at $t\to\infty$  and so the equation (\ref{24}) reduces to the Born rule in momentum space $g(p)=|\tilde\phi(p)|^2$  \cite{Feynman 1965, Holland}. But finding an explicit expression for momentum probability density in the relativistic regime is not straightforward, therefore a numerical calculation of $g(p)$ for the Gaussian wave packet (\ref{19}) is presented in Figure \ref{kkk}. This numerical study indicates that the relativistic momentum probability density deviates significantly from Born rule only when the width of the wave function in momentum space is greater than $mc$.

\section{Conclusion and outlook} 
In this paper, in a simple case of single free spinless particle in $(1+1)$-dimensions, we have extracted a "reasonable" probability density current. By "reasonable" we mean: the current   i) is manifestly covariant, ii) is conserved,  iii) has a non-negative first component, iv) does not lead to faster than light particle propagation and v) reduces to Born probability current density in non-relativistic limit. These conditions naturally give rise to the given probability density current. Therefore, at least in $(1+1)$-dimensions, probabilistic interpretation of relativistic spinless wave function is possible. Extending this study to  $(3+1)$-dimensional interacting particle systems will be the next step.
Such systems should be described by quantum filed theory. The state of a system in quantum filed theory is an arbitrary vector in the appropriate Fock space and may well involve a superposition
of states of different particle numbers, namely $\ket{\Psi}= \sum_n \int  {\tilde{\phi}}_n(p_1,..,p_n) \ket{p_1,..,p_n}$.
It evolves according to the appropriate Schr\"{o}dinger equation;  $i \partial_t\ket{\Psi}=\text{H} \ket{\Psi} $ where $\text{H}$ is the Hamiltonian operator in Schr\"{o}dinger picture. In the presence of interaction this equation leads to  a system of coupled integro-differential equations for multi-particle wave functions, $\phi_n $, a reccurrent procedure in the literature of Light-front quantization \cite{Brodsky 1998}. In future works, we aim to find a probabilistic interpretation for these wave equations in position space.
\section{Acknowledgement}
We thank M.M. Sheikh-Jabbari, Y. Rokni and H. Abedi for helpful discussions and  A. Deriglazov for carefully reading the manuscript and his useful comments.
\section{Appendix A}
In this appendix, we will derive the equations (\ref{8}), (\ref{12}) and (\ref{13}) in $(1+1)$-dimensions . Without loss of generality, the wave function can be expanded as a linear combination of plane waves:
\begin{equation}\label{25}
	\phi(x)=\sum_{n} A_n e^{ip_n.x}.
\end{equation}	
Plugging this into (\ref{6}) and using (\ref{7}) yields,
\begin{eqnarray}\label{26}
\begin{split}
J^0\pm J^1=\sum_{n,m} A_n A_m^* \alpha(p_n,p_m) (p^\pm_n + p^\pm_m) e^{i(p_n-p_m).x},
\end{split}
\end{eqnarray}
where $p^\pm_n=p^0_n \pm p^1_n$. In $(1+1)$-dimensions, conditions $ J^\mu J_\mu\geq0$ and $J^0\geq0$ lead to 
\begin{equation}\label{27}
J^0\pm J^1\geq 0,
\end{equation}
for arbitrary wave-functions. Therefore we can consider 
\begin{equation}\label{28}
	\alpha(p_n,p_m)= \frac{[F^\pm(p_n)] \ [F^\pm(p_m)]^*+[F^\pm(p_n)]^* \ [F^\pm(p_m)]}{p^\pm_n+p^\pm_m},
\end{equation}
which leads to the following positive definite expression for $J^0\pm J^1$: 
\begin{equation}\label{29}
J^0\pm J^1= \left|\sum_n F^\pm(p_n) A_n e^{ip_n.x}\right|^2+\left|\sum_n [F^\pm(p_n)]^* A_n e^{ip_n.x}\right|^2, 
\end{equation}
where $F^\pm(p_n)$ is an unknown function which must be determined. Since the only scalar can be made by $p_n$ is the rest mass, a dimensional analysis leads to $|\alpha(p_n,p_n)|=\frac{1}{2m_0}$; the factor $1/2$ is a convention and can be absorbed in normalization constant. 
Therefore,
\begin{equation}\label{30}
F^\pm(p_n)=e^{i\lambda^{\pm}(p_n)}\sqrt{{p^\pm_n}/{2m_0}}. 
\end{equation}
Whether one substitutes $F^+$ or $F^-$ the resulted $\alpha$ is the same. This fact can be used to determine phase of $F^\pm$ as $\lambda^\pm(p_n)=\pm l \pi$, where $l$ is an integer number. Then we have
\begin{equation}\label{31}
	\alpha(p_n,p_m)= \frac{[\sqrt{p^\pm_n}] \ [\sqrt{p^\pm_m}]^*+[\sqrt{p^\pm_n}]^* \ [\sqrt{p^\pm_m}]}{2m_0(p^\pm_n+p^\pm_m)}.
\end{equation} 
 A straightforward but tedious calculation shows that, $\alpha(p_n,p_m)$ can be rewritten as equation (\ref{8}) which ensures that $\alpha(p_n,p_m)$ is a scalar. Also from equations (\ref{29}) and (\ref{30}), it is clear that 
\begin{equation}\label{32}
\begin{split}
J^0=  |\sum_n \sqrt{\frac{p^+_n}{4m_0}} A_n e^{ip_n.x}|^2+|\sum_n [\sqrt{\frac{p^+_n}{4m_0}}]^* A_n e^{ip_n.x}|^2  \\
 +|\sum_n \sqrt{\frac{p^-_n}{4m_0}} A_n e^{ip_n.x}|^2 +|\sum_n [\sqrt{\frac{p^-_n}{4m_0}}]^* A_n e^{ip_n.x}|^2,
 \end{split}
\end{equation}
\begin{equation}\label{33}
\begin{split}
J^1=  |\sum_n \sqrt{\frac{p^+_n}{4m_0}} A_n e^{ip_n.x}|^2+|\sum_n [\sqrt{\frac{p^+_n}{4m_0}}]^* A_n e^{ip_n.x}|^2  \\
 -|\sum_n \sqrt{\frac{p^-_n}{4m_0}} A_n e^{ip_n.x}|^2 -|\sum_n [\sqrt{\frac{p^-_n}{4m_0}}]^* A_n e^{ip_n.x}|^2.
 \end{split}
\end{equation}
Finally, using the definition of ${\mathcal{D}^{\pm}}$ operators, (\ref{14}), equations (\ref{32}) and (\ref{33}) reduce to (\ref{12}) and (\ref{13}).

\section{Appendix B}
In this appendix, we will show there is a sequence of positive energy wave functions, $\{\phi_n\}_{n=1}^{\infty}$, whose corresponding probability density sequence, $\{\rho_n\}_{n=1}^{\infty}$, approaches $\delta(x-a)$ as a generalized function \cite{Lighthill}. This argument is closely similar to that of Bracken and  Melloy \cite{Bracken 1999} for the case of Dirac electron.
 
Consider following  sequence of positive energy wave functions:
\begin{equation}\label{B1}
\phi_n(x)=\int \sqrt{\frac{m}{n E(p)}}f(\frac{p}{n})e^{ip(x-a)}  dp,
\end{equation}
in which $E_p=\sqrt{p^2+m^2}$ and $f(p)$ is  a normalized Gaussian function,  $\int |f(p)|^2dp=1$, defined as:
 \begin{equation}\label{B2}
f(p)=(1/{m\sqrt{\pi}})^{\frac{1}{2}} e^{-p^2/2m^2}.
\end{equation}
Substituting Eq.(\ref{B1}) in Eq.(\ref{16}), reads
\begin{equation}\label{B3}
\begin{split}
\rho_n(x)=\frac{1}{n}\left|\int S_+(p)f(p)e^{ip(x-a)}dp\right|^2 \\+\frac{1}{n}  \left|\int S_-(p)f(p)e^{ip(x-a)} dp\right|^2 ,
\end{split}
\end{equation}
where $S_{\pm}=\sqrt{\frac{E_p\pm p}{2 E_p}}$. Using convolution theorem, the Fourier transform of probability density, $\tilde{\rho}_n(p)$, can be written as
\begin{equation}\label{B4}
 \tilde{\rho}_n(p)=R_n(p)\frac{e^{-ipa}}{\sqrt{2\pi}},
\end{equation}
in which
\begin{equation}\label{B5}
R_n(p)=\frac{1}{n}\int f(\frac{q-p}{n})f(\frac{q}{n}) \Gamma(q-p,q) dq,
\end{equation}
\begin{equation}\label{B6}
\Gamma(q,p)=S_+(q)S_+(p)+S_-(q)S_-(p).
\end{equation}
Since the Fourier transform of $\delta(x-a)$  is  $\frac{e^{-ipa}}{\sqrt{2\pi}}$, we need to show that $\lim_{n\to\infty} R_n(p)=1$. For this , we consider $q=nr$ and rewrite $R_n(p)$ as 
\begin{equation}\label{B7}
R_n(p)=\int f(r-\frac{p}{n})f(r) \Gamma(nr-p,nr) dr.
\end{equation}
A straightforward calculation shows that $\Gamma(q,p)$ can be rewritten as
\begin{equation}\label{B8}
\Gamma(q,p)=G_1(q)G_1(p)+G_2(q)G_2(p),
\end{equation}
in which
\begin{equation}\label{B9}
G_1(p)=\sqrt{\frac{E_p+m}{2E_p}},
\end{equation}
\begin{equation}\label{B9}
G_2(p)=\frac{p}{\sqrt{2E_p(E_p+m)}}.
\end{equation}
By Taylor's Theorem, we have
\begin{equation}\label{B10}
f(r-\frac{p}{n})=f(r)-\frac{p}{n} f'(r-\eta\frac{p}{n}),
\end{equation}
\begin{equation}\label{B11}
G_i(nr-p)=G_i(nr)-\frac{p}{n} G_i'(nr-\theta p),
\end{equation}
where $0\leq\eta\leq 1$ and $0\leq\theta\leq 1$. Using equations (\ref{B10}) and (\ref{B11}), the Eq.(\ref{B7}) reads
\begin{equation}\label{B12}
R_n(p)=A_n(p)+B_n(p)+C_n(p)+D_n(p),
\end{equation}
where
\begin{equation}\label{B13}
A_n(p)=\int |f(r)|^2 \Gamma(nr,nr)dr,
\end{equation}
\begin{equation}\label{B14}
B_n(p)=-\frac{p}{n}\int f'(r-\eta \frac{p}{n})f(r) \Gamma(nr,nr)dr,
\end{equation}
\begin{equation}\label{B15}
C_n(p)=\frac{p^2}{n}\int  f'(r-\eta\frac{p}{n})f(r) \Upsilon(nr-\theta p,nr) dr,
\end{equation}
\begin{equation}\label{B16}
D_n(p)=-p\int |f(r)|^2  \Upsilon(nr-\theta p,nr) dr,
\end{equation}
\begin{equation}\label{B17}
\Upsilon(q,p)= G_1'(q)G_1(p)+ G_2'(q)G_2(p).
\end{equation}
Finally, after tedious calculations we get
\begin{equation}\label{B18}
\begin{aligned}
&A_n(p)=1,
\\
\lim_{n\to\infty} B_n(p)=\lim_{n\to\infty} &C_n(p)=\lim_{n\to\infty} D_n(p)=0,
\end{aligned}
\end{equation}
which show the probability density sequence, $\{ \rho_n(x)\}$, converges to $\delta(x-a)$ as $n$ tends to infinity.
 

\end{document}